\documentclass[twocolumn,resetfootnote,twocolappendix,longbib]{aastex7}

\usepackage[utf8]{inputenc}
\usepackage{tipa}
\usepackage{combelow}
\usepackage{savesym}
\savesymbol{tablenum}
\usepackage{siunitx}
\restoresymbol{SIX}{tablenum}
\usepackage{breqn}
\usepackage{tabularx}
\usepackage{placeins}

\newcommand{\sssb}[3]{#1 #2$_{#3}$}
\newcommand{\occult}{\texttt{occult3d}}
\newcommand{\uk}{G!kún\textdoublepipe’hòmdímà}
\newcommand{\uksat}{G!ò’é~!hú}
\newcommand{\ukboth}{G!kún\textdoublepipe’hòmdímà--G!ò’é !hú}
\newcommand\multimoon{\texttt{MultiMoon}}

\begin{document}

\title{Triaxial shapes and densities of \uk, Haumea, and Varda from stellar occultations}

\author[orcid=0000-0002-1788-870X,sname='Proudfoot']{Benjamin Proudfoot}
\affiliation{Florida Space Institute, University of Central Florida, 12354 Research Parkway, Orlando, FL 32826, USA}
\email[show]{benp175@gmail.com}

\author[orcid=0000-0002-8296-6540, sname='Grundy']{Will Grundy} 
\affiliation{Lowell Observatory, 1400 W Mars Hill Rd, Flagstaff, AZ 86001, USA}
\affiliation{Northern Arizona University, Department of Astronomy \& Planetary Science, PO Box 6010, Flagstaff, AZ 86011, USA}
\email{grundy@lowell.edu}

\author[orcid=0000-0002-6085-3182]{Flavia Luane Rommel}
\affiliation{Florida Space Institute, University of Central Florida, 12354 Research Parkway, Orlando, FL 32826, USA}
\email{}

\author[orcid=0000-0003-2132-7769]{Estela Fernández-Valenzuela}
\affiliation{Florida Space Institute, University of Central Florida, 12354 Research Parkway, Orlando, FL 32826, USA}
\email{estela@ucf.edu}

\author[orcid=0000-0003-1080-9770, sname='Ragozzine']{Darin Ragozzine} 
\affiliation{Brigham Young University Department of Physics \& Astronomy, N283 ESC, Brigham Young University, Provo, UT 84602, USA}
\email{darin_ragozzine@byu.edu}


\begin{abstract}
The shapes and densities of mid-sized and large trans-Neptunian objects (TNOs) are pivotal for understanding a variety of important aspects of planet formation. In this work, we present a Bayesian shape modeling method which combines constraints from rotational light curves and satellite orbits to construct three-dimensional shape models of TNOs. We use it to reanalyze three stellar occultations of the TNOs (229762) \uk{} (2007 UK$_{126}$), (136108) Haumea, and (174567) Varda. By assuming that their satellites (or ring) orbit in their respective equatorial planes, we are able to derive unique shape models for both \uk{} and Haumea. Our derived shape for \uk{} is \added{spheroidal} with $a = b = 329^{+4}_{-3}$ km and $c = 294^{+11}_{-10}$ km, with a system density $\rho = 1007^{+50}_{-49}$ kg m$^{-3}$. For Haumea, we find $a = 1061^{+87}_{-71}$ km, $b = 844^{+5}_{-7}$ km, and $c = 514^{+18}_{-19}$ km, providing $\rho = 2050^{+157}_{-152}$ kg m$^{-3}$. For Varda, after updating its mutual orbit with its satellite Ilmarë, we find that currently published data are unable to fully constrain its three-dimensional shape. Intriguingly, Varda's elongated limb appears to point towards its satellite at the time of the occultation. With a $\sim$2\% chance of such an alignment happening randomly, this may be \added{suggestive of} a frozen-in tidal and/or rotational bulge. Our work emphasizes the importance of how external constraints can improve occultation analyses. With continued observations of rotational light curves, stellar occultations, and satellite orbits, these---and other---TNOs can have their shapes and densities further refined.

\end{abstract}

%


\section{Introduction} 
\label{sec:intro}

Due to their large heliocentric distances, \added{relatively little} is known about the sizes, shapes, and densities of trans-Neptunian objects (TNOs). As the largest remnants of the process of planet formation---apart from the known planets---these bodies preserve clues to the conditions in the protoplanetary disk and the processes that governed planetary growth \citep{morbidelli2020kuiper}. Understanding the shapes and densities of TNOs is key to understanding the formation and evolution of TNOs \citep{mckinnon2008structure}. 

Shape provides interesting constraints on formation \citep{leinhardt2010formation,nelsen2025beyond}, collisional alteration \citep{rommel2023large}, and hydrostatic relaxation \citep{ortiz2017size}, while density can help to infer composition \citep{grundy2019mutual}, porosity \citep{brown2013density}, and thermal history \citep{desch2009thermal}. Measurements of bulk density also provide critical inputs into models of a TNO's internal structure \citep{guilbert2020internal}, which can inform the study of potential subsurface oceans \citep{hussmann2006subsurface} and ability to generate/retain volatile species \citep{schaller2007volatile}. The ensemble of shape and density measurements across the known TNOs also provides tight constraints on the timing and/or formation mechanisms of TNOs \citep{bierson2019using}. 

Despite their importance, high-precision size determinations remain available for only a small fraction of the known TNO population. With angular sizes of tens of mas (or less), TNO diameters are $\sim$unable to be directly measured from ground- or space-based telescopes, except in a few cases \citep[e.g.][]{stern1997hst,buie2010pluto}. Thermal radiometry with space-based observatories such as Spitzer and Herschel \citep{lellouch2013tnos,muller2020trans} has been successful in estimating effective diameters, but the resulting uncertainties can often be too imprecise to estimate densities \citep{brown2017density}, while also requiring assumptions about the (often unknown) shape and spin-axis orientation \citep{kiss2019gonggong}. On the other hand, mass measurements can be obtained fairly straightforwardly for any TNOs with a satellite/binary companion \citep[e.g.][]{grundy2019mutual}. But for singleton objects, mass is currently impossible to measure \citep{2025AJ....170..353F}.

The shapes of TNOs, which affect density measurements, are even more poorly constrained. Rotational light curves (RLCs) can constrain elongation and rotational properties \citep{sheppard2008photometric,fernandez2022modeling}, but provide non-unique solutions, are affected by degeneracies between shape and albedo variegation, and require assumptions about surface scattering properties. Recently, work has been done to constrain the shapes of TNO binaries based on non-Keplerian orbital precession \citep{proudfoot2024bpm2}, but this is only available for a few TNOs so far \citep{proudfoot2024beyond, nelsen2025beyond}.

\added{Shape is a critical input into understanding the role of hydrostatic equilibrium in TNOs \citep{ortiz2020stellar}. Generally, large TNOs have been assumed to be in (or at least near) hydrostatic equilibrium, which predict Maclaurin spheroid or Jacobi ellipsoid shapes if TNOs act like self-gravitating fluids \citep{tancredi2008dwarfs}. The role of material strength and inter-particle friction also play a role in the shapes of TNOs \citep{2007Icar..187..500H}. Similarly, the effects of non-homogeneous density structures, which are expected in large differentiated TNOs, can also alter the shape of a TNO \citep{rambaux2017equilibrium,dunham2019haumea}. However, with such few measured three-dimensional shapes, little can be said about the relative roles of each of these effects. Thus, further determinations of TNO shapes are necessary to better understand the physical characteristics of these bodies.}

One of the best techniques to understand both size and shape simultaneously is stellar occultations \citep{ortiz2020stellar}. Over the past few decades, with the Gaia stellar catalog \citep{vallenari2023gaia} and TNO ephemerides \citep{desmars2015orbit}, occultations of many TNOs have now been observed. Unfortunately, occultations only provide an instantaneous view of the limb of a TNO, with further work required to derive a true three-dimensional shape and size. Multiple occultations, RLC constraints, and orientation constraints from rings and/or satellites can help to provide these more complex shape models.

In this work, we develop a new software tool to fit occultation chords to triaxial shape models constrained by satellite orbit poles and RLCs. We present our new software tool, named \occult, in Section \ref{sec:occult3d}. Then, we apply it to \uk, Haumea, and Varda in Sections \ref{sec:uk126}, \ref{sec:haumea}, and \ref{sec:varda}, where we derive triaxial shapes and densities. We then discuss population details and conclude in Section \ref{sec:discussion}.

\section{\occult}
\label{sec:occult3d}

Stellar occultations allow the measurement of the instantaneous elliptical limb of a TNO. While useful, one measurement of the limb alone cannot uniquely determine the three-dimensional shape of a TNO. With additional occultations, information derived from RLCs, or orientation constraints based on ring or satellite geometry, degeneracies can be broken providing a unique shape model. To allow for simultaneous use of all of these constraints, we have created \occult, a publicly available Python-based software.

\occult{} derives three-dimensional shape models of TNOs by casting the occultation fitting problem as a Bayesian parameter inference exercise. At its core, occultation chords (from an arbitrary number of events) are fit to three-dimensional shape models, while information about the target's orientation and RLC amplitude can be included as prior probabilities. Instead of $\chi^2$ optimization or similar frequentist statistical methods, \occult{} explores the model parameter space using a Markov Chain Monte Carlo (MCMC) approach. For simplicity, we choose the \texttt{emcee} sampler, an ensemble MCMC sampler that is widely used across a variety of fields \citep{foreman2013emcee}. This allows \occult{} to flexibly explore the entire parameter space and allows for easy identification of model degeneracies. With limited occultations, RLC knowledge, or other information, three-dimensional shape models can have significant degeneracies that provide non-unique shape models. 

The forward model at the core of \occult{} has 8 free parameters (when analyzing a single occultation). Six of these parameters determine the three-dimensional shape (triaxial semi-axes $a,b,c$), orientation (pole direction $\alpha,\delta$), and rotational phase ($\phi$) of the target. Two additional terms ($f_c,g_c$) provide the ephemeris offset of the limb's center on the plane of the sky. Based on the shape/orientation parameters, the elliptical limb profile of the target can be derived using the equations of \citet{magnusson1986distribution}. The $\chi^2$ of the occultation chords (compared to the elliptical limb) can then be found by radial limb fitting \citep[described in detail in, e.g, ][]{sicardy2011pluto,ortiz2012albedo}.


Assuming independent, normally distributed errors, the Bayesian log-likelihood ($\mathcal{L}$) is proportional to $\frac{1}{2}\chi^2$. Priors can then be included to further refine shape models \citep[e.g.,][]{brown2013size}. In this version of \occult, we focus on priors based on RLC knowledge and orientation constraints from rings and satellites. 

First, we focus on RLC-derived constraints. Assuming that the RLC of a body is determined by the variation in projected limb area alone (i.e., neglecting albedo variegation and realistic scattering properties), the RLC amplitude of a triaxial TNO is given by the equation:
\begin{equation}\label{eqn:rlc}
  \Delta m = -\frac{5}{2} \log\left[\left(\frac{b}{a}\right)\left(\frac{(a/c)^2\cos^2\theta+\sin^2\theta}{(b/c)^2\cos^2\theta + \sin^2\theta}\right)^{1/2}\right]
\end{equation}
\noindent where $a$, $b$, $c$ are the body's ellipsoidal semi-axes and $\theta$ is the polar aspect angle. A polar aspect angle of 0$\degr$ (90$\degr$) corresponds to pole-on (equator-on) geometry. Therefore, with a known RLC amplitude, significant constraints on shape models can be found \citep[e.g.,][]{ortiz2017size}. When imposed in \occult, RLC amplitude priors are normally distributed, taking an amplitude ($\Delta m$) as the mean and the (1$\sigma$) amplitude uncertainty ($\sigma_{\Delta m}$) as the scale. \added{When light curve measurements are taken at different times than occultations, some evolution of a body's aspect angle is expected, leading to slow changes in the RLC amplitude \citep[e.g.,][]{fernandez2019changing}. To account for this, we apply the light curve prior with the aspect angle of the body at the time of the light curve observations, not the occultation. This allows our modeling to be as robust as possible to secular changes in the light curves of these bodies.}

This formulation assumes that the entire RLC is caused by variations in the projected limb area, with no contributions from albedo variegation across a body's surface. Typically the surfaces of small bodies have uniform albedos, validating this assumption, although famous counter-examples do exist. Pluto notably has a high-amplitude RLC \citep[$\Delta m \sim 0.3$,][]{tholen1994pluto} despite a spherical shape. Likewise, Haumea is known to have a dark red spot on its surface, although the extent of this feature is currently unconstrained \citep{lacerda2008high}. 

\added{Equation \ref{eqn:rlc} also implicitly assumes the scattering behavior of the surfaces. Differences in scattering behavior and albedo variegation can radically change an RLC, biasing any inferred three-dimensional shape output. To account for these issues, judicious and parsimonious selection of priors is necessary. Although in this work we assume the formalism provided by Equation \ref{eqn:rlc}, where important, we revisit these assumptions and explore how altering them could lead to different shape models. In particular, our discussion of Haumea in Section \ref{sec:haumea} provides a detailed examination of how scattering and albedo affect final shape models.}

Although RLC amplitudes alone can narrow the range of allowable shape models, the constraints they place are limited without knowledge of the rotational phase at the time of occultation ($\phi$). For some bodies with a high amplitude, well-measured RLC, it is possible to infer the rotational phase at the time of the occultation, but more often, the rotational phase is unknown. This is mostly because rotational periods are not known with the precision required to calculate the rotational phase at the moment of the occultation, unless a RLC has been taken relatively close in time to the occultation event. Thankfully, with the Bayesian approach used by \occult, $\phi$ can be left as a free parameter, allowing the MCMC sampler to explore different values. In the future, with more numerous occultations, it will be possible to use a rotational period and single phase (for a given reference epoch) to eliminate these free parameters. This will be particularly important for cases like Quaoar, which has been observed during stellar occultations many times \citep[e.g.,][]{margoti2024quaoarshape}.

Lastly, \occult{} can include constraints on the orientation of the target. For some bodies, like Haumea, the presence of a ring naturally provides an independent measure of the body's pole orientation. Rings like those around Haumea and Quaoar, should lie very close to their parent's equatorial plane as any inclination will naturally be damped by differential precession caused by the TNO's non-spherical shape \citep{tiscareno2014planetary,marzari2020ring}. Although less definitive, the presence of a satellite can also probe a body's orientation. For many large TNOs, satellites are expected to lie in (or near) their parent's equatorial plane \citep{sicardy2024stellar}. Confirmation of this alignment has yet to be made for most TNO-satellite systems, but where the alignment is independently measured, they tend to be well-aligned \citep{brozovic2015orbits,proudfoot2024beyond,braga2025investigating}. These constraints are further included as priors in \occult, where the prior on the pole RA/dec ($\alpha,\delta$) is assumed to be normally distributed (see orbit fit in Section \ref{sec:varda} and Appendix \ref{sec:appendix_orbit})\footnote{$\alpha,\delta$ are given by $\alpha=\Omega-90\degr$ and $\delta=90\degr - i$, where $i,\Omega$ are the inclination and longitude of the ascending node, in the equatorial frame.}. 

\added{Although the dominant shape of large TNOs is expected to be overall triaxial, topography on the surface of TNOs (e.g., craters, mountains, etc.) can cause significant departures from an idealized triaxial shape \citep[see, for example,][]{rommel2023large}. This can be especially important with chords with ultra-precise timing uncertainties. So as to not bias the triaxial shape model output by \occult, an additional model uncertainty can be added to account for topography ($\sigma_{\rm topo}$). This uncertainty, which can be set arbitrarily based on physical expectations \citep[e.g.,][]{1973Icar...18..612J}, is added in quadrature to the chord extremity uncertainties based on timing, similar to implementations in other works \citep{Morgado2021,SORAcitation,rommel2023large}. We note that in this work, we use $\sigma_{\rm topo}=0$ km, as our shape models generally provide good fit quality without it. }

\added{To measure the goodness-of-fit for each of our \occult{} fits, we use the mean normalized squared residuals, defined as 
\begin{align}
    {\rm MNSR}=\frac{1}{N}\sum_i \left( \frac{y_{\rm obs} - y_{\rm model}}{\sigma_{obs}} \right)^2
\end{align}
\noindent where $N$ is the number of chord extremities, $y_{\rm obs},y_{\rm model}$ are the observed and modeled chord extremities, and $\sigma_{\rm total}$ is the total uncertainty in the chord extremity (i.e., including the user-specified topographic uncertainty). We use this in place of more traditional goodness-of-fit metric (like reduced $\chi^2$) as the number of effective degrees-of-freedom in our model is not well-defined given the presence of parameters that are strongly bounded by priors and/or strongly degenerate. The MNSR is equivalent to the average $\chi^2$ of each chord extremity. For high-quality fits, we expect $\rm{MNSR}\sim1$, while poor-quality fits will have $\rm{MNSR}\gg1$. With $\rm{MNSR}\ll1$, uncertainties in the chord extremities are likely to be overestimated. }

For simplicity, much of \occult{} uses the functionality provided by the \textbf{S}tellar \textbf{O}ccultation \textbf{R}eduction and \textbf{A}nalysis (SORA) package \citep{SORAcitation}. The tools provided by SORA are used for translating occultation chords to the sky plane, evaluating the $\chi^2$ of trial limb profiles, and filtering solutions based on close negative chords. These tools are well-validated and used across a variety of stellar occultation work \citep[e.g.][]{rommel2023large,Kretlow2024,rommel2025stellar,Rizos2025}. We point the reader to \citet{SORAcitation} for a detailed explanation of occultation analysis with the SORA library.

\added{\occult{} has been made publicly available on GitHub\footnote{\url{https://github.com/benp175/occult3d}}, alongside a citable Zenodo \dataset[DOI]{https://doi.org/10.5281/zenodo.20288143}. There, full installation instructions and a quick start guide are provided. We also provide a full working example that reproduces the analysis we present in Section \ref{sec:uk126}. We intend to add additional functionality to \occult{} in the future, allowing for more complex photometric modeling, additional shape models, automated hydrostatic equilibrium analysis, fitting to multiple occultations, among others. We welcome contributions from the occultation community and hope that \occult{} will enhance analysis of TNO occultations.}

\begin{deluxetable*}{lcccc}
\tablecaption{Triaxial shape fitting results}
\tablehead{
 & \uk{} Triaxial & \uk{} Maclaurin & Haumea & Varda (best fit)
}
\startdata
Best fit MNSR & 1.31 & 1.36 & 1.41 & 0.86 \\
\hline
\multicolumn{5}{c}{\textbf{Priors}} \\
$a$ (km) & $0 < a < 5000$ & $0 < a < 5000$ & $0 < a < 5000$ & $0 < a < 5000$ \\
$b$ (km) & $0 < b < a$ & $0 < b < a$ & $0 < b < a$ & $0 < b < a$ \\
$c$ (km) & $0 < c < b$ & $0 < c < b$ & $0 < c < b$ & $0 < c < b$ \\
$\alpha$ ($\degr$)  & $20.6\pm1.5$      & $20.6\pm1.5$      & $285.1\pm0.5$ & $272.6\pm1.5$        \\
$\delta$ ($\degr$)  & $46.25\pm0.32$    & $46.25\pm0.32$    & $-10.6\pm1.2$ & $-10.8\pm2.0$        \\
$\phi$ ($\degr$)    & \nodata           & \nodata           & $75.6\pm3.6$ & \nodata               \\
$f_{\rm c}$ (km)     & \nodata          & \nodata          & \nodata & \nodata             \\
$g_{\rm c}$ (km)     & \nodata          & \nodata   & \nodata & \nodata             \\
$\Delta m$ (mag) & $0.03\pm0.01$     & \nodata            & $0.32\pm0.10$ & \nodata              \\
\hline
\multicolumn{5}{c}{\textbf{Fitted Parameters}} \\
$a$ (km)       & $336^{+8}_{-6}$   & $329^{+4}_{-3}$   & $1061^{+87}_{-71}$ & 389             \\
$b$ (km)       & $325^{+5}_{-7}$   & \nodata   & $844^{+5}_{-7}$ & 353                \\
$c$ (km)       & $295^{+11}_{-11}$ & $294^{+11}_{-10}$ & $514^{+18}_{-19}$ & 248              \\
$f_{\rm c}$ (km)     & $11^{+4}_{-4}$          & $11^{+4}_{+4}$          & $-130^{+14}_{-13}$ & -20             \\
$g_{\rm c}$ (km)     & $23^{+7}_{-7}$          & $-21^{+6}_{-7}$   & $-174^{+9}_{-10}$ & -128             \\
\hline
\multicolumn{5}{c}{\textbf{Derived Parameters}} \\
$r_{\rm vol}$ (km) & $318^{+5}_{-5}$   & $317^{+5}_{-4}$   & $772^{+20}_{-19}$ & 324              \\
$\rho$ (kg m$^{-3}$)    & $997^{+58}_{-54}$ & $1007^{+50}_{-49}$ & $2050^{+157}_{-152}$ & 1726                  \\
$c/a$ & $0.88^{+0.04}_{-0.04}$ & $0.89^{+0.04}_{-0.03}$ & $0.49^{+0.04}_{-0.04}$ & 0.64  \\
$b/a$ & $0.96^{+0.01}_{-0.01}$ & \nodata & $0.79^{+0.06}_{-0.06}$ & 0.91  \\
\enddata
\tablecomments{Fitting for Varda does not return a unique shape model, so we show the best fit model found, although we point out that the $c$-axis is $\sim$uncorrelated with goodness-of-fit. Both pole orientation angles are referenced to the J2000 equatorial coordinate system and are defined as the direction of the spin angular momentum in a right-handed coordinate system. Ephemeris centers are referenced to the NIMAv11, JPL\#125, and NIMAv14 ephemerides for \uk, Haumea, and Varda, respectively \citep[for more information on NIMA, see][]{desmars2015orbit}. Priors are implemented as Gaussian distributions with mean and standard deviation as given. Where no data is given for priors, priors are not implemented. For fitted parameters, no data are given for quantities which are not applicable (e.g., $b$-axis in the Maclaurin model since $b=a$, by definition). Light curve priors take into account any change in aspect angle between the light curve measurements and the occultation. Density ($\rho$) is the system density for \uk{} and Varda, but for Haumea, it is solely Haumea's density (i.e. no satellite contribution).}
\label{tab:results}
\end{deluxetable*}

\begin{figure*}
    \centering
    \includegraphics[width=\textwidth]{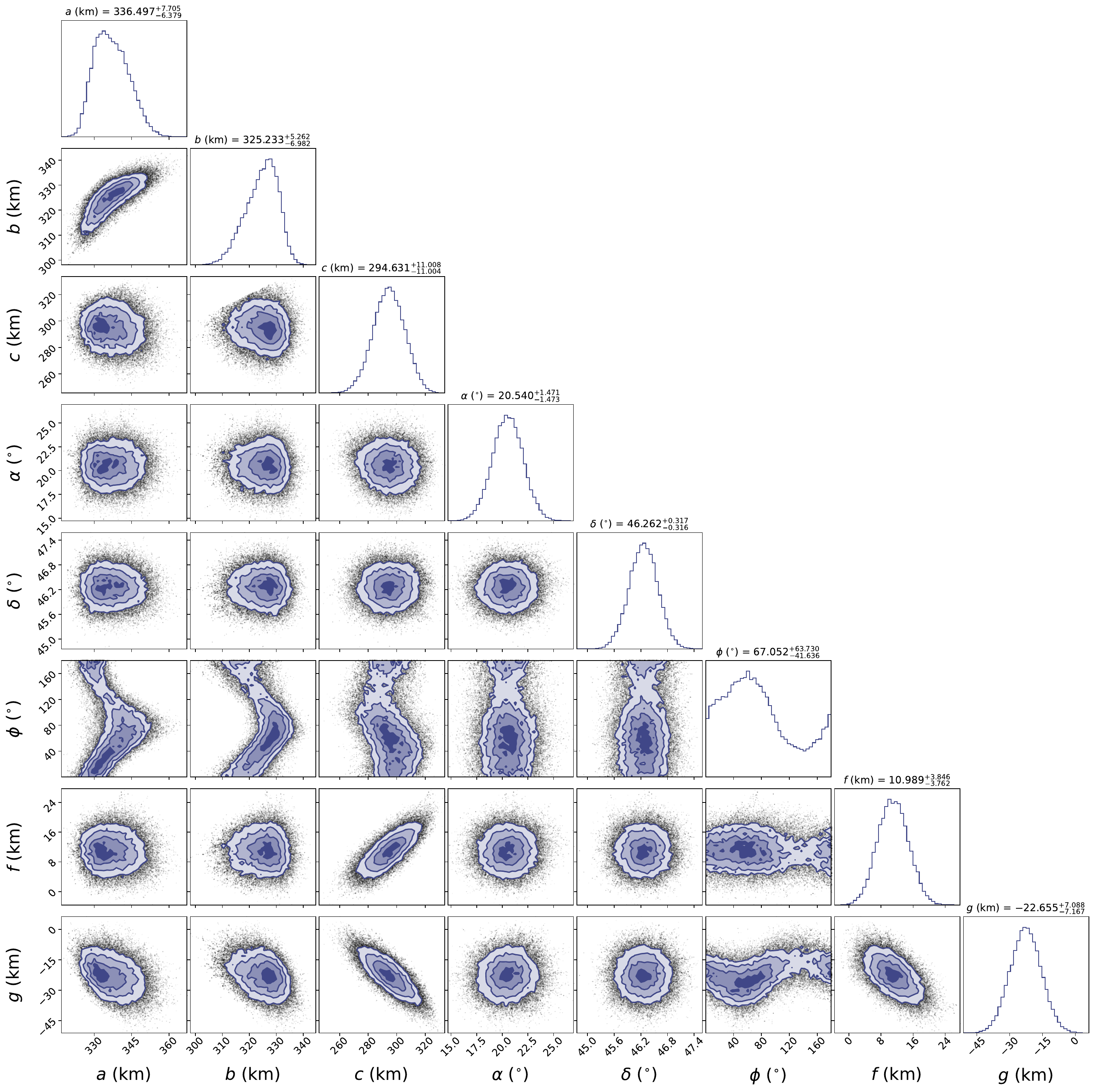}
    \caption{A corner plot showing the triaxial shape model derived for \uk. Marginal (one-dimensional) parameter posterior distributions are shown along the tops of each column, while joint (two-dimensional) parameter distributions for each pair of parameters are shown as contour plots. Contours show the 0.5, 1, 1.5, and 2 $\sigma$ confidence intervals. Black points show individual samples from the MCMC chain. }
    \label{fig:corner_uk126}
\end{figure*}

\section{G!kún\textdoublepipe'hòmdímà}
\label{sec:uk126}

\uk{} (2007 UK$_{126}$) was observed during an occultation in November 2014. A total of 8 chords were obtained during the occultation campaigns, providing a well-constrained elliptical limb profile \citep{benedetti2016results,schindler2017results}. Observations of \uk's RLC found a low amplitude $\Delta m = 0.03\pm0.01$ \citep[\added{at average epoch 2011-10-30,}][]{thirouin2014rotational}. In addition, \uk{} has a satellite---\uksat---which can place a strong constraint on the orientation of \uk{} if we assume it is on an equatorial orbit \citep{grundy2019mutual}. 

\begin{figure}
    \centering
    \includegraphics[width=\columnwidth]{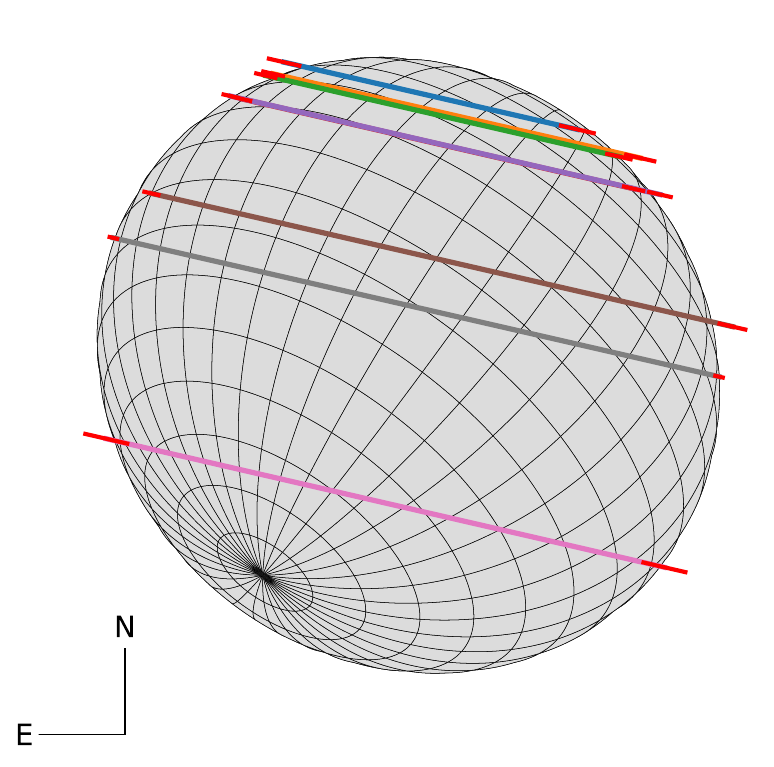}
    \caption{Our best-fit triaxial shape model for \uk. Colored lines show the various occultation chords detected during the 2014 stellar occultation \citep{benedetti2016results,schindler2017results}, while the red line tips show the uncertainty in the start and end of the occultation chords. This best fit model corresponds to a shape of $a = 339$ km, $b = 326$ km, and $c = 298$ km. }
    \label{fig:uk126_3d}
\end{figure}

Using \occult, we fit a three-dimensional shape model to \uk's occultation chords. These chords are compiled in Table \ref{tab:chords}, along with all chords for subsequent analyses. The results of our shape fitting are displayed in Table \ref{tab:results}, along with the MNSR measuring the goodness-of-fit. We also show the posterior for our analysis as a corner plot in Figure \ref{fig:corner_uk126}. With such a low amplitude RLC, the resulting shape model is nearly azimuthally symmetric ($a\approx b$) with $a = 336^{+8}_{-6}$ km, $b = 325^{+5}_{-7}$ km, and $c = 295^{+11}_{-11}$ km. This yields a volumetric radius (the radius of a sphere with identical volume) of $r_{vol} = 318^{+5}_{-5}$ km. We compare the three-dimensional shape model and occultation chords in Figure \ref{fig:uk126_3d}.

Using the system mass of $(136.1\pm3.3) \times10^{18}$ kg \citep{grundy2019mutual}, we can also derive the system density. Based on a size ratio of $4.45\pm0.08$ between \uk{} and \uksat{} \citep{grundy2019mutual}, our triaxial shape yields a density of $\rho = 997^{+58}_{-54}$ kg m$^{-3}$ for the system, when assuming that \uksat{} has an equivalent albedo, shape, and density.

Our derived shape model is very close to a \added{spheroidal} ($a/b= 1$) due to the low RLC amplitude and low subobserver geometry. Therefore, we should also consider that, with such a small RLC amplitude, the photometric variations may be due to albedo variegations on the surface of \uk, rather than a triaxial shape. In this case, \occult{} gives $a = b = 329^{+4}_{-3}$ km and $c = 294^{+11}_{-10}$ km ($c/a = 0.89^{+0.04}_{-0.03}$) and a density of $\rho = 1007^{+50}_{-49}$ kg m$^{-3}$. This model provides a similar likelihood as the triaxial model. 

Although using equilibrium figures is an imperfect way to understand the shapes of TNOs \citep[see][]{2007Icar..187..500H}, their use provides a physically-motivated model that can provide some guidance in data-limited regimes. To understand whether \added{the spheroidal shape of \uk{} is consistent with being a Maclaurin spheroid in hydrostatic equilibrium}, we can compare our shape model to that expected of a fluid at a given rotation and density. Using the Chandrasekhar formalism \citep{chandrasekhar1987ellipsoidal}, a Maclaurin spheroid will satisfy the following equation:
\begin{equation}\label{eqn:maclaurin}
    \frac{\Omega^2}{\pi G \rho} = \frac{2\sqrt{1-e^2}}{e^3}\left( 3-2e^2\right)\arcsin{e} - \frac{6}{e^2}\left( 1-e^2 \right)
\end{equation}
\noindent where $\Omega$ is the angular velocity, $G$ is the gravitational constant, $\rho$ is the density, and $e^2 = 1- c^2/a^2$. Numerically solving this equation for $\Omega$, given our measured shape and density, we find \uk{} is in/near hydrostatic equilibrium when/if its rotational period is between 9.8 and 14.5 hours. This is very close to the 11.05 h period reported in the literature, although other possible period aliases exist \citep{thirouin2014rotational}. \added{This provides compelling evidence---although not conclusive given the uncertainty in rotation period---that} if \uksat{} orbits in \uk's equatorial plane, \uk{} is likely in (or nearly in) hydrostatic equilibrium. \added{Alternatively, it could retain }a hydrostatic equilibrium shape that was frozen-in at an earlier epoch (potentially even at formation). \added{Further confirmation of the shape model and better RLC periods can provide more insight into this possibility.}

The success of our modeling approach shows that even a single occultation can help to refine the triaxial shape of mid-sized TNOs when combined with RLCs and reasonable assumptions about the orientation of satellite orbits. 

Confirming our assumptions about \uk's orientation will require substantial dedication of observations resources. Although RLCs can help to distinguish orientation solutions \citep[e.g.,][]{Tegler2005Period,Fernandez2017Physical,fernandez2019changing}, the low-amplitude RLC of \uk{} and slow sky movement make this process nearly impossible. In addition, if the light curve is due to albedo variegation, the RLC amplitude changes due to changing aspect angle can be complex, especially if albedo features are localized. 

In future occultations, our \added{spheroidal} shape model predicts that the limb shape will be similar to that previously observed, with only small changes due to \uk's heliocentric motion. Another observation of a similar limb shape can help to confirm our shape model, but cannot independently measure the pole orientation. 

One route forward is to observe the precession of \uksat's orbit. Nodal precession can reveal the obliquity of the primary (with respect to the satellite orbit), but cannot provide a unique pole orientation \citep{proudfoot2024bpm2}. Given the shape we derive, we expect that the nodal precession rate is $\sim1-3\degr$ yr$^{-1}$, with a total precession period of $\sim100-300$ years. This may be detectable with new astrometric measurements of the system as the earliest observations of the system date to 2008. With a known shape, even a non-detection of nodal precession can provide strong constraints on the pole orientation of \uk.

We encourage continued observations of \ukboth{} which will further refine the system's density, shapes, mutual orbits, and orientations. 

The inferred bulk density for \ukboth{}---$\rho = 1007^{+50}_{-49}$ kg m$^{-3}$---is quite low for its size. Assuming it has a similar composition to other large TNOs (typically with densities $\sim1800$ kg m$^{-3}$), such a low density requires a porosity of $\sim$45\% \citep[for further discussion, see][]{grundy2019mutual}. This seems to stand in contrast to its fairly regular (near) hydrostatic equilibrium shape, as the same physics that produces such equilibrium shapes, gravity overcoming viscosity/strength, will tend to compress pores. The presence of a small presumably collisionally-formed moon, like those around other large TNOs \citep{barr2016interpreting}, also would suggest an extensive collisional history which should provide some compaction of the upper 10s of km of the surface \citep{milbury2015crater,bierson2019using}. 

The case of \uk{} is similar to that of Uni (provisionally designated \sssb{2002}{UX}{25}), which has a small satellite, similar size, and low bulk density---although its density is only inferred from thermal measurements \citep{brown2013density,brown2017density}. Just larger than this size range, densities appear to rapidly increase (see Section \ref{sec:discussion}), a transition that has proved difficult to explain \citep{bierson2019using,loveless2022structure,canas2024solution}. Perhaps more complex models, which may simultaneously account for pore space collapse, collisions, internal melting, possible compositional differences \citep[like those proposed by][]{canas2024solution}, and/or other complex geophysical mechanisms, will provide more insight into the formation of these fascinating transitional bodies.

\section{Haumea}
\label{sec:haumea}
In 2017, a multi-chord occultation of Haumea was captured by almost a dozen telescopes over Europe \citep{ortiz2017size}. From this event, a three-dimensional shape of Haumea was successfully derived based on the orientation of Haumea's rings and its light curve, assuming that Haumea was at its light curve minimum. However, as pointed out by \citet{dunham2019haumea}, the assumption of minimum rotational phase can significantly change the implied shape of Haumea, and may not have been fully justified. Although Haumea was certainly near its minimum, photometry from around the time of the occultation shows that occultation occurred briefly before rotational minimum \citep[see Extended Data Figure 6 in][]{ortiz2017size}. Here, we conduct a full reanalysis of the occultation data using \occult.

To constrain shape models of Haumea, we perform a very similar analysis as \citet{ortiz2017size}, but instead allow the rotational phase to vary around its expected value. Based on Extended Data Figure 6 from \citet{ortiz2017size}, which shows the RLC of Haumea phased to the time of occultation, the occultation occurred at a phase of $0.04\pm0.01$ (or $14.4\pm3.6\degr$) before minimum. Hence, in our \occult{} fits, we place a prior of $\phi = 90\degr - (14.4\pm 3.6)\degr$, where $90\degr$ corresponds to the minimum. 

As previously done, we leverage the orientation of Haumea's ring to constrain Haumea's pole orientation. Rings around oblate bodies like Haumea quickly have any inclination damped, which minimizes differential precession that can increase collisional activity \added{\citep{tiscareno2014planetary,marzari2020ring}}. Haumea's satellites are too far away to significantly perturb ring particles, and play practically no role in the orientation of the rings, making this a very safe assumption \citep[e.g.,][]{marzari2020ring}. Hence, we place priors of $\alpha = 285.1\pm 0.5\degr$ and $\delta = -10.6\pm 1.2\degr$ to match the ring orientation found in \citet{ortiz2017size}.

Haumea's RLC amplitude has been well-studied over the decades since its discovery \citep[e.g.][]{lacerda2008high,lockwood2014size}. Importantly, \citet{lockwood2014size} studied Haumea's RLC with the Hubble Space Telescope (HST), which provided resolved photometry of the Haumea system and was able to resolve Haumea's RLC without dilution from its satellites. This RLC has an amplitude of $\Delta m = 0.32$ \added{(on 2009-02-04)}, however, it is unclear how much of the amplitude is from shape and/or albedo. Indeed, the two RLC minima have a difference of $\sim$0.05 mag. 

In addition to this, Equation \ref{eqn:rlc} cannot account for more realistic surface properties; \citet{lockwood2014size} points out that when using a more realistic surface, the required axes ratios are less extreme than would be expected from Equation \ref{eqn:rlc} alone. For example, they suggest a $b/a = 0.80\pm0.01$ using a photometric model based on Uranus' moon Ariel to match $\Delta m = 0.32\pm0.006$ mag. In comparison, equation \ref{eqn:rlc} gives $\Delta m = 0.24\pm0.01$ mag for those axes ratios (when using the same aspect angle). This suggests that uncertainties of $\sim0.1$ mag on the RLC amplitude could be present. Ideally, we could instead use a realistic photometric model to more accurately model the photometric behavior, but we defer this to future work. 

Alternatively, RLC constraints could be eliminated altogether by fitting multiple occultations simultaneously, but no such data are publicly available. To remain as conservative as possible in our shape modeling and account for both the uncertainties in the shape/albedo degeneracy and the lack of a realistic photometric model, we place a prior of $\Delta m = 0.32\pm 0.10$ mag. This may be overly pessimistic, but it more precisely accounts for our lack of knowledge about Haumea and its RLC. When using this prior, we find that the occultation chords provide a slightly better fit when assuming a lower RLC amplitude around $\sim0.24$ mag, suggesting that $\Delta m = 0.32$ mag may indeed be too large (when assuming Eqn. \ref{eqn:rlc}).

\begin{figure}
    \centering
    \includegraphics[width=\columnwidth]{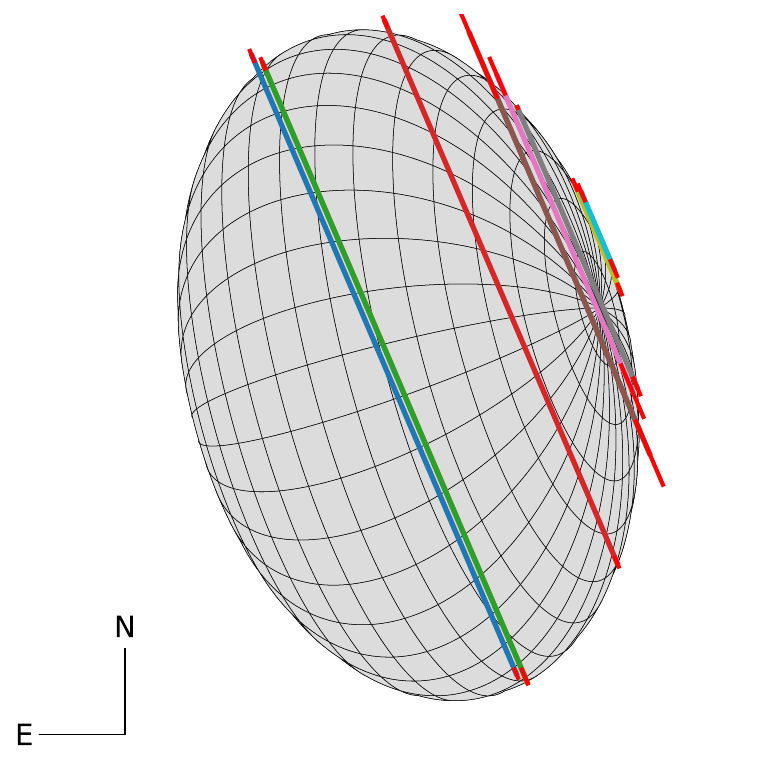}
    \caption{Best-fit shape model of Haumea, in the style of Figure \ref{fig:uk126_3d}. The shape model shown here corresponds to a triaxial shape with $a = 1072$ km, $b = 842$ km, and $c = 515$ km. Occultation chords are those given in \citet{ortiz2017size}.}
    \label{fig:haumea_3d}
\end{figure}

Using all these constraints, we fit a triaxial model to Haumea based on the 2017 occultation chords (all chords are presented in Table \ref{tab:chords}). Our results are shown in detail in Table \ref{tab:results}. We found a triaxial shape with $a = 1061^{+87}_{-71}$ km, $b = 844^{+5}_{-7}$ km, and $c = 514^{+18}_{-19}$ km. We plot our best-fit model in Figure \ref{fig:haumea_3d}. This yields a volume-equivalent radius of $r_{vol} = 772^{+20}_{-19}$ km. Combined with the most recent measurement of Haumea's mass, $(3952\pm11)\times10^{18}$ kg \citep[][]{proudfoot2024beyond}, this yields a total bulk density of Haumea of $2050^{+157}_{-152}$ kg m$^{-3}$.

\added{Compared to \citet{ortiz2017size} ($a = 1161\pm30$ km, $b=852\pm4$ km, $c=513\pm16$ km, $r_{vol}=798\pm6$ km, and $\rho=1857\pm42$ kg m$^{-3}$ when using the updated mass), our solution is broadly similar in overall triaxial geometry but does show systematic differences in the inferred axis lengths. These differences are primarily driven by two methodological choices, our treatment of rotational phase as a free parameter and differences in the adopted photometric constraints and uncertainty estimates. One of \occult's strengths is its ability to fully propagate uncertainties associated with model assumptions (e.g., rotational phase or spin-axis geometry), resulting in more realistic parameter uncertainties in the present analysis.}

The triaxial shape we derive roughly matches the model suggested by hydrostatic equilibrium of a two-layer differentiated model of Haumea \citep{dunham2019haumea}, albeit with a slightly smaller $c$-axis. If borne out by future occultations, this smaller $c$-axis could indicate that Haumea is slightly out of hydrostatic equilibrium. Alternatively, a more complicated internal model with additional layers---possibly a subsurface ocean---could provide a better match to Haumea's shape. 

Unfortunately, Haumea is in a sparse star field, making occultations relatively rare occurrences. However, even just a few positive occultation chords will enable far better constraints on Haumea's triaxial shape, and may enable analyses without assumptions about Haumea's RLC, sidestepping any issues stemming from inaccurate photometric models. Future occultations will also provide an opportunity to further refine the ring orientation, further improving shape modeling efforts.


\begin{deluxetable*}{cccCCCC}
\tablewidth{\textwidth}
\tablecaption{Observed Astrometric Positions of Ilmarë}
\tablehead{
Julian Date & Date & Telescope/Instrument & \Delta \alpha \cos{\delta} & \sigma_{\Delta \alpha \cos{\delta}} & \Delta \delta & \sigma_{\Delta \delta} \\
 & & & ('') & ('') & ('') & ('')
}
\startdata
2454947.91380 & 2009-04-26 & HST/WFPC2   & +0.12311  & 0.00334 & -0.01033 & 0.00201 \\
2455411.77394 & 2010-08-03 & Keck/NIRC2  & -0.09206 & 0.00200 & -0.10862 & 0.00200 \\
2455411.84777 & 2010-08-03 & Keck/NIRC2  & -0.08012 & 0.00200 & -0.11149 & 0.00200 \\
2455439.72274 & 2010-08-31 & HST/WFC3    & -0.13093 & 0.00097 & -0.00172 & 0.00246 \\
2455441.04295 & 2010-09-01 & HST/WFC3    & -0.02228 & 0.00131 & -0.13318 & 0.00137 \\
2455467.09059 & 2010-09-27 & HST/WFC3    & -0.00562 & 0.00467 & +0.13799  & 0.00172 \\
2455752.29017 & 2011-07-09 & HST/WFC3    & +0.08235  & 0.00858 & -0.11167 & 0.00413 \\
2456020.03911 & 2012-04-02 & Gemini/NIRI & -0.11187 & 0.00300 & +0.07594  & 0.00300 \\
2456053.96807 & 2012-05-06 & Gemini/NIRI & -0.05069 & 0.00300 & +0.12971  & 0.00300 \\
2456141.85712 & 2012-08-02 & Gemini/NIRI & -0.10084 & 0.01354 & -0.07780 & 0.00618 \\
2456404.98806 & 2013-04-22 & Gemini/NIRI & -0.08404 & 0.00300 & +0.09796  & 0.01080 \\
2456486.75571 & 2013-07-13 & Gemini/NIRI & -0.12196 & 0.00332 & -0.08097 & 0.00786 \\
2459453.80128 & 2021-08-27 & Keck/NIRC2  & -0.14815 & 0.00300 & -0.00494 & 0.00300 \\
2460537.83575 & 2024-08-15 & Keck/NIRC2  & +0.15285  & 0.00300 & +0.01362  & 0.00300 \\
2460575.74686 & 2024-09-22 & Keck/NIRC2  & -0.12305 & 0.00391 & -0.07866 & 0.00300 \\
\enddata
\tablecomments{Observations from 2013 and before are taken verbatim from \citet{grundy2015mutual}.}
\label{tab:observations}
\end{deluxetable*}

\section{Varda}
\label{sec:varda}

\subsection{Updating the mutual orbit}
Here, we provide an updated orbit fit for the Varda-Ilmarë binary system. Although an orbit solution was derived in \citet{grundy2015mutual}, that work provided two mirror-ambiguous orbit solutions with different orbit pole directions. With enough time since the last set of observations, the breaking of this mirror ambiguity is now possible.

To do this, we acquired three Keck observations of Varda and Ilmarë from 2021-2024 using the laser guide star adaptive optics system \citep{wizinowich2006} with the NIRC2 camera\footnote{\url{https://www2.keck.hawaii.edu/inst/nirc2}}. Observations were taken in the infrared $H$ filter, with wavelengths between $\sim$1.48 to 1.77 $\mu$m, and were dithered to allow for sky-subtraction. Astrometry was extracted using well-validated methods described in the literature \citep[e.g.][]{grundy2015mutual}. In addition to these new observations, we use the available relative astrometry in the literature \citep{grundy2015mutual}; our entire dataset is shown in Table \ref{tab:observations}.

Using these new observations, orbit fitting was able to rule out the retrograde orbit at $6\sigma$ confidence. With a single orbit solution, we performed a more detailed orbit fit using \multimoon, a Bayesian orbit fitter designed for fitting TNO binary orbits \citep[][]{ragozzine2024beyond}. We used the Keplerian orbit fitting module, see \citet{ragozzine2024beyond} and \citet{proudfoot2024bpm2} for further description of how \multimoon{} functions. Our fits were run with 960 walkers for 45,000 total steps (20,000 burn-in, 5,000 post-pruning burn-in, 20,000 sampling). Convergence of the fits was assessed based on posterior smoothness, best-fit sample quality, and inspection of walker trace plots. The orbit residuals and posterior distribution are shown in Appendix \ref{sec:appendix_orbit}.

Our final orbit solution is shown in Table \ref{tab:fits}. Although slightly different from the past orbit solution, this is not unexpected given the new data. Interestingly, our best-fit orbit solution has a $\chi^2_\nu$ ($\chi^2$ per degree of freedom) of $\sim$1.8. This confirms previous findings that Varda-Ilmarë's mutual orbit appears to have a significant non-Keplerian component, possibly due to the non-spherical shape of Varda (or Ilmarë) \citep{proudfoot2024bpm2}. We leave further analysis of the non-Keplerian component of the orbit to future work.

\begin{deluxetable*}{lcc}
\tabletypesize{\footnotesize}
\tablecaption{Keplerian Orbit Solution for Varda-Ilmarë}
\tablehead{
Parameter &  & Posterior
}
\startdata
\textbf{\textit{Fitted parameters}}  &               &                                 \\
System mass ($10^{18}$ kg)           & $M_{\rm sys}$     & $267.5^{+4.9}_{-4.8}$           \\
Semi-major axis (km)                 & $a$           & $4815^{+29}_{-29}$              \\
Eccentricity                         & $e$           & $0.016^{+0.004}_{-0.004}$       \\
Inclination ($\degr$)                & $i$           & $77.4^{+1.9}_{-1.9}$            \\
Argument of periapsis ($\degr$)           & $\omega$      & $307^{+15}_{-19}$                \\
Longitude of the ascending node ($\degr$)            & $\Omega$      & $2.6^{+1.5}_{-1.5}$           \\
Mean anomaly at epoch ($\degr$)               & $\mathcal{M}$ & $145^{+19}_{-15}$               \\
\hline
\textbf{\textit{Derived parameters}} &               &                                 \\
Orbit period (d)         & $P_{\rm orb}$     & $5.750824^{+0.000016}_{-0.000016}$ \\
Orbit pole RA ($\degr$)              & $\alpha_{\rm orb}$  & $272.6^{+1.4}_{-1.5}$                         \\
Orbit pole dec. ($\degr$)            & $\delta_{\rm orb}$  & $-10.8^{+2.0}_{-1.9}$                         \\
\enddata
\tablecomments{Reported values represent the median value and uncertainties are based on 16th and 84th percentiles. All fitted angles are relative to the J2000 ecliptic plane on Varda-centric JD 2455300 (2010 Apr. 14 12:00 UT), except for RA and dec. values which are referenced to the J2000 equatorial coordinate system. }
\vspace{-1.15cm}
\label{tab:fits}
\end{deluxetable*}

\begin{figure}
    \centering
    \includegraphics[width=\columnwidth]{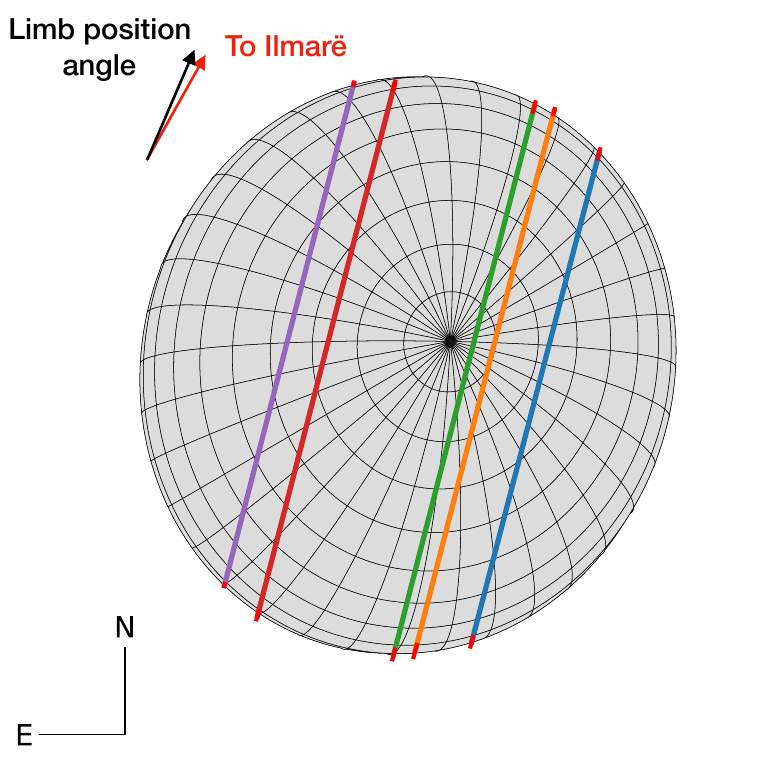}
    \caption{Best-fit shape model for Varda, in the style of Figure \ref{fig:uk126_3d}. The shape model shown here corresponds to a triaxial shape with $a = 389$ km, $b = 353$ km, and $c = 248$ km. We point out that a wide range of shape models---especially those with a wide range of $c/a$ ratios---are allowable given the current constraints. The red arrow points towards the direction of Ilmarë during the occultation, while the black arrow points to the best fit position angle (although has an uncertainty of $8\degr$). Occultation chords are those given in \citet{souami2020multi}. }
    \label{fig:varda_3d}
\end{figure}

\subsection{Occultation fitting}

In 2018, Varda was observed during a stellar occultation over the USA. Five chords from this event have been published \citep{souami2020multi}, with an additional 15 positive chords reported in a conference abstract, but have not yet been published \citep{schindler2019results}. Even with the five publicly available chords, valuable constraints can be placed on Varda's size and shape. Combining these occultation chords with constraints from Ilmarë's orbit pole and Varda's RLC, a possible triaxial shape model can be derived.

First looking to priors from Varda's RLC, \citet{thirouin2014rotational} used four years worth of photometric data in an attempt to recover Varda's RLC period. The best fit provided a \added{single-peaked} rotation period of 5.91 hours and amplitude of $0.02\pm0.01$ mag \added{(at average epoch 2011-07-15)}, but with many similar peaks in the periodogram, confidence of that period is relatively low. Other analyses showed \added{similar periods (and aliases) but with higher RLC amplitudes of $0.06\pm0.02$ mag \citep{thirouin2010short}. }
Given the relatively large size ratio (comparable to Pluto-Charon), tidal synchronization should happen relatively quickly, especially at the small semi-major axis of the binary \citep{thirouin2014rotational}. Salacia-Actaea, another similarly sized TNO binary, was recently discovered to be tidally synchronized despite early indications that showed a rapid $\sim$6.5 hour RLC period \citep{collyer2025synchronous}. With no definitive rotation solution and a pole-on orbit of Ilmarë, we do not place any constraints on either the RLC amplitude or the rotational phase. 

We place a prior on Varda's pole orientation to match the orientation of Ilmarë's orbit (see Table \ref{tab:fits}). In the case of a tidally evolved Varda and Ilmarë, this assumption is well-justified as it is the tidal end state of any binary system \citep{hut1980}. Even if not fully tidally evolved, alignment of the orbit and rotational axes is likely \citep{sicardy2024stellar}.

Using \occult, we fit a triaxial shape model to Varda. Although our best fit solution (shown in Table \ref{tab:results}) provides a reasonable shape model with $a = 389$ km, $b = 353$ km, and $c = 248$ km (Figure \ref{fig:varda_3d}), a wide range of shape models are allowable (hence why we only provide the best fit values). Good-fit shape models typically have $b/a \sim 0.9$, although a non-unity value is favored only at $\sim1.5\sigma$. Our models, however, have unconstrained values for $c/a$. This is due to Varda's nearly pole-on orientation (i.e. we view Varda along the $c$-axis) making it difficult to infer any information along the line-of-sight. A pole-on geometry is indeed consistent with the low apparent photometric variability \citep{thirouin2014rotational}.

Based on the reported occultation chords, \citet{souami2020multi} provided a detailed analysis of Varda's shape under the assumption of a Maclaurin spheroidal shape. They found that a Maclaurin shape seemed to be consistent with the occultation limb if one of the mirror orbit solutions was chosen. Unfortunately, this orbit solution has now been conclusively ruled out. If Varda is indeed aligned with Ilmarë's orbit, Varda's apparent elongation ($a>b$) suggests a triaxial, rather than spheroidal, shape. If rotating rapidly, Varda's triaxial shape could be due to rotational deformation. 

\begin{figure*}
    \centering
    \includegraphics[width=0.9\textwidth]{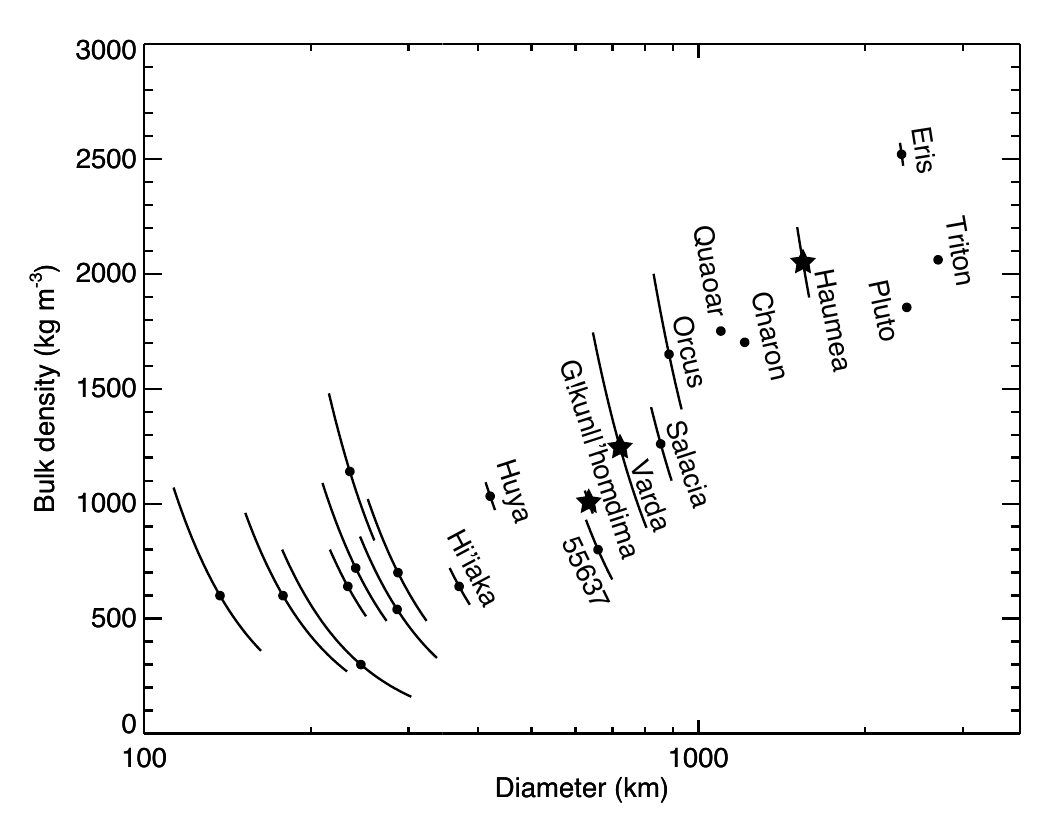}
    \caption{The size-density relationship of TNOs. Sizes/densities updated here are shown by stars. For Varda, we use previous size/density measurements \citep{grundy2015mutual} along with our new system mass. Other densities are from \citet{proudfoot2025beyond}, \citet{rommel2025stellar}, \citet{grundy2019mutual}, \citet{2025NatCo..1610926F}, and references therein. Unlabeled data points from lower left to upper right are: Typhon, Teharonhiawako, Altjira, Ceto, Sila, Lempo, (612239), and (26308). }
    \label{fig:density}
\end{figure*}

A rotating Jacobi ellipsoid in hydrostatic equilibrium will satisfy the following relationship:
\begin{align}\label{eqn:jacobi}
\begin{split}
    \frac{a^2b^2}{b^2-a^2} \left[ f(a^2,b^2,c^2,a^2) - f(a^2,b^2,c^2,b^2) \right] = \\ c^2f(a^2,b^2,c^2,c^2)
\end{split}
\end{align}
\noindent where $a,b,c$ are the semi-axes of the ellipsoid and $f$ is the elliptic integral:
\begin{equation}
    f(x,y,z,p) = \frac{3}{2}\int^{\infty}_{0}\frac{dt}{(t+p)\sqrt{(t+x)(t+y)(t+z)}}
\end{equation}
\noindent \citep{darwin1887iii,olver2010nist} \added{where $t$ is the quantity being integrated over and $p$ is an arbitrary input}. Numerically solving this equation using our best fit $a = 389$ km and $b = 353$ km, we find $c = 216$ km. This yields a system density of 1900 kg m$^{-3}$, when using a diameter ratio between Varda and Ilmarë of 1.95 \citep{grundy2015mutual} and the system mass from Table \ref{tab:fits}. A Jacobi ellipsoid satisfies:
\begin{align}\label{eqn:jacobispin}
\begin{split}
    \frac{\omega^2}{\pi G \rho} =
    \frac{4abc}{3(a^2-b^2)} [ a^2f(a^2,b^2,c^2,a^2) - \\ b^2f(a^2,b^2,c^2,b^2)]
\end{split}
\end{align}
\noindent which allows calculation of the rotation period required for such a Jacobi ellipsoid. We find our best fit would need to rotate with a period of $\sim4.5$ hours to be explained by rotational deformation. \added{With such an extreme shape, the light curve would likely be dominated by the tri-axial shape, as such the single-peaked light curve period would be half this value. This period is much shorter than that derived by \citet{thirouin2014rotational}, and the density implied appears somewhat too high for an object the size of Varda (see Figure \ref{fig:density})}. We note that our shape models provide a range of allowable $a$ and $b$ values (and therefore hydrostatic $c$ values), so solutions with reasonable density \added{and rotation period} may exist. 

Tidal deformation due to Ilmarë can also alter Varda's shape. The equilibrium figure of a binary (assuming fluid bodies) in synchronous rotation has its surface defined by a Roche ellipsoid \citep{roche1850,chandrasekhar1963equilibrium}. Based on the sequences of Roche ellipsoids tabulated by \citet{Leone1984}, with a mass ratio $M_2/M_1\approx0.08$, rotation/orbital period of 5.75 days, and density of $\sim1500$ kg m$^{-3}$, we estimate Varda's current tidal deformation at $b/a > 0.99$---a trivial amount. As such, we can confidently rule out tidal forces at their present strength as the source of Varda's possible triaxial shape.

Alternatively, Varda could have a non-hydrostatic equilibrium shape. We note that, at the time of the 2018 occultation, Varda's long ($a$) axis appears to point towards Ilmarë (see Figure \ref{fig:varda_3d}). The projected two-dimensional elliptical limb of Varda has a position angle of $67\pm8\degr$ \citep{souami2020multi}, while Ilmarë is at a position angle of $60.8\pm0.7\degr$. The probability of such a close alignment (or better) happening at random---if Varda is non-synchronously rotating---is $\sim$2\%. Such an alignment is the minimum energy state expected of a fully tidally evolved system. If Varda is rotating synchronously, as we suggest, the ellipticity of Varda's limb could represent a frozen-in shape (i.e., a fossil bulge) from a time when it had a much more rapid rotation (allowing it to take on a triaxial shape) and/or a more pronounced tidal bulge. \added{We do note that non-fluid shapes are also possible when granular physics, internal friction, non-homogeneous density distributions, and other factors are taken into account \citep{2007Icar..187..500H,rambaux2017equilibrium}. A full accounting of these effects requires a better shape model for Varda.}

Additional RLC monitoring and observations of occultations will provide further clarification on Varda's possible triaxial shape and fossil bulge. Just a single multi-chord occultation with another apparent alignment would provide strong evidence for (or against) a fossil bulge. Similarly, additional occultation chords from the 2018 event may show Varda's elliptical limb in more detail, and could probe the edges of the limb where more information about the $c$-axis is available (see Figure \ref{fig:varda_3d}).

Given the lack of a unique shape model, a precise measure of Varda's volumetric radius or density cannot be derived. With a pole-on geometry---if Varda is indeed aligned with Ilmarë's orbital plane---it will be difficult to probe Varda's $c$-axis. Although idealized models---like Jacobi ellipsoids---can provide theoretical constraints, they rely on a well-measured RLCs and assume hydrostatic equilibrium. Thus, it seems unlikely that \textit{empirical} shape models for Varda will improve considerably without either remote observations \citep[like the long-range observations conducted by \textit{New Horizons},][]{verbiscer2022diverse}, until Varda's heliocentric motion brings it to a more favorable viewing geometry. If not tidally relaxed, non-Keplerian shape effects may also provide interesting constraints on Varda's shape, although any constraints will require assumptions about Varda's interior structure \citep{proudfoot2024beyond}.

\section{Discussion and Conclusions}
\label{sec:discussion}
The densities of mid-sized and large TNOs is of utmost importance for understanding the formation, composition, and evolution of TNOs \citep{grundy2019mutual,canas2024solution}. Indeed, several works have used the distribution of TNO densities to infer the composition and timing of TNO formation \citep[e.g.,][]{barr2016interpreting,bierson2019using}. Here, we have shown that by performing a holistic analysis of TNO occultations, light curves, and satellite orbits, we can provide more precise measurements of the density of TNOs. We show these updated densities in Figure \ref{fig:density}. 

The most obvious interpretation of density variation is distinct compositional differences between objects, where planetary bodies born in certain regions of the protosolar disk may have inherited different initial inventories of ices and refractory materials \citep{brown2013density}. One interesting model, which uses both icy and rocky pebbles in a streaming instability and pebble accretion simulation, found that larger planetesimals that undergo pebble accretion will become rockier over time \citep{canas2024solution}. The model posits that smaller pebbles are more likely to have their ices depleted by UV irradiation, and these pebbles are preferrentially incorporated into TNOs via pebble accretion, and provide a good match to the TNO size-density relationship. 

In contrast, many evolutionary processes are also likely to shape the densities of TNOs. Collapse of pore space appears to be an important process \citep{mckinnon2008structure,bierson2019using}. Differentiation and internal melting can speed this process, allowing liquid to fill macroscopic porosity \citep{mckinnon2008structure}. On the other hand, collisions may also be able to strip outer, less dense layers of a TNO leaving a more dense remnant \citep[e.g.][]{barr2016interpreting}.

With a variety of competing models, no agreement has yet been reached on the exact cause of the size-density relation of TNOs. The solution, however, is likely to be multi-faceted, with a variety of complex formational and evolutionary processes contributing. Hopefully, more comprehensive models that combine many of these processes can further illuminate the source of the density dichotomy. Likewise, an improvement in the density determinations of TNOs will enable higher confidence in density trends \citep{lyra2025missing}.

Past a broader look at the ensemble of TNO densities, individual densities are also important for understanding the interiors of TNOs. Densities allow a rough determination of the icy and rocky inventories, and therefore sources of long-term radiogenic heating \citep{parker2021review}. Larger, denser TNOs have far greater inventories of radionuclides, allowing for more extensive melting and a greater potential for subsurface oceans \citep{desch2009thermal}. These bodies have also probably undergone either partial or full differentiation, allowing for formation of a dense rocky core and a water-rich mantle. In these bodies, geochemical processes may be active, producing the methane rich surfaces seen on Pluto, Eris, and Makemake \citep{grundy2024measurement,glein2024moderate}. Smaller, less dense TNOs like \uk{} likely have never had widespread subsurface melting and remain largely undifferentiated bodies \citep{grundy2019mutual}. 

Determining shapes is also critical for understanding TNOs. Most clearly, without a precise determination of the shape of a TNO, its density remains poorly constrained, as in the case of Varda (see Section \ref{sec:varda}). Beyond that, however, shape is an important measurement for a variety of phenomena. Most notably, a secure determination of hydrostatic equilibrium requires precise knowledge of a TNO's shape \citep{tancredi2008dwarfs}. Shape can also reveal the inner structure of a TNO, as hydrostatic equilibrium shapes are different for non-homogenous interiors \citep{rambaux2017equilibrium,dunham2019haumea}. Relatively little is known about the shapes of the largest TNOs, and with increasing numbers of precise stellar occultation observations, we will begin to illuminate the shape distribution of TNOs.

We note that there are significant difficulties in assuming simple hydrostatic equilibrium shapes of TNOs \citep{2007Icar..187..500H,dunham2019haumea}. Although Jacobi and Maclaurin shapes provide a starting point towards understanding TNO shapes, various complications like viscosity, strength, friction, and non-homogeneous interiors make them \textit{imperfect} shape models. With the explosion of occultation science, we are quickly approaching a regime where simple models should be abandoned for empirically determined shape models and realistic theoretical models motivated by reasonable geophysical parameters \citep{dunham2019haumea}.


To help ascertain both triaxial shapes and densities, we have developed \occult, which is able to flexibly explore shape models of TNOs. The flexibility of a Bayesian parameter inference approach allows \occult{} to flexibly incorporate a variety of physical constraints, which we have shown can vastly improve triaxial shape models of TNOs. To provide this ability to the community, we have made \occult{} publicly available, and plan on various upgrades as occultation science progresses. These upgrade will add functionality to fit multiple occultations, focus on hydrostatic equilibrium figures, and add more realistic surface photometric behavior. 

We encourage ongoing observations of TNOs, whether long-term photometric monitoring, occultation observations, or continued tracking of TNO satellites. Holistic analyses using all these observations can provide in-depth views into these distant, icy worlds.


\begin{acknowledgments}
We thank Bryan Holler for helpful discussions that supported this work. We thank two anonymous reviewers for their feedback that improved the readability of the manuscript. We also thank the BYU Office of Research Computing for their dedication to providing computing resources without which this work would not have been possible. Additionally, we thank the numerous occultation observers involved in originally observing these occultations.

The authors wish to recognize and acknowledge the very significant cultural role and reverence that the summit of Maunakea has always had within the Native Hawaiian community. We are most fortunate to have the opportunity to conduct observations from this mountain. 

Some of the data presented herein were obtained at Keck Observatory, which is a private 501(c)3 non-profit organization operated as a scientific partnership among the California Institute of Technology, the University of California, and the National Aeronautics and Space Administration. The Observatory was made possible by the generous financial support of the W. M. Keck Foundation. 

B.P. and F.L.R. acknowledge the generous support of the UCF Preeminent Postdoctoral Program (P$^3$).

\end{acknowledgments}

\begin{contribution}

B.P. was responsible for development, testing, and using \occult, and was primarily responsible for writing the text of the article. W.G. conducted Keck observations of Varda-Ilmarë and conducted preliminary orbit fitting for that system. F.L.R. gave extensive feedback on occultation limb fitting. D.R. provided access to computational resources for the orbit fitting. E.F.-V. supervised B.P. and provided training on understanding occultations. All authors provided editing, commentary, and feedback on the manuscript.


\end{contribution}

\bibliography{all}
\bibliographystyle{aasjournalv7}



\appendix
\restartappendixnumbering

\section{Orbit Fit Diagnostic Plots}
\label{sec:appendix_orbit}
\begin{figure}[b]
    \centering
    \includegraphics[width=\columnwidth]{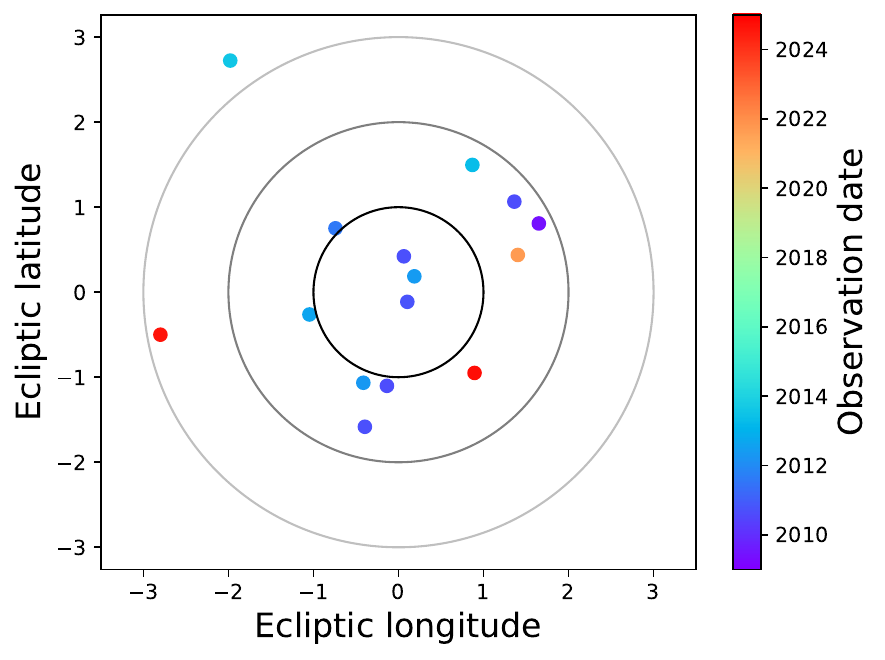}
    \caption{Residuals for the Varda-Ilmarë orbit fit. Circles show the 1, 2, and 3$\sigma$ contours. The $\chi^2$ per degree of freedom for this fit is $\sim$1.8. }
    \label{fig:residual}
\end{figure}

\begin{figure*}
    \centering
    \includegraphics[width=\textwidth]{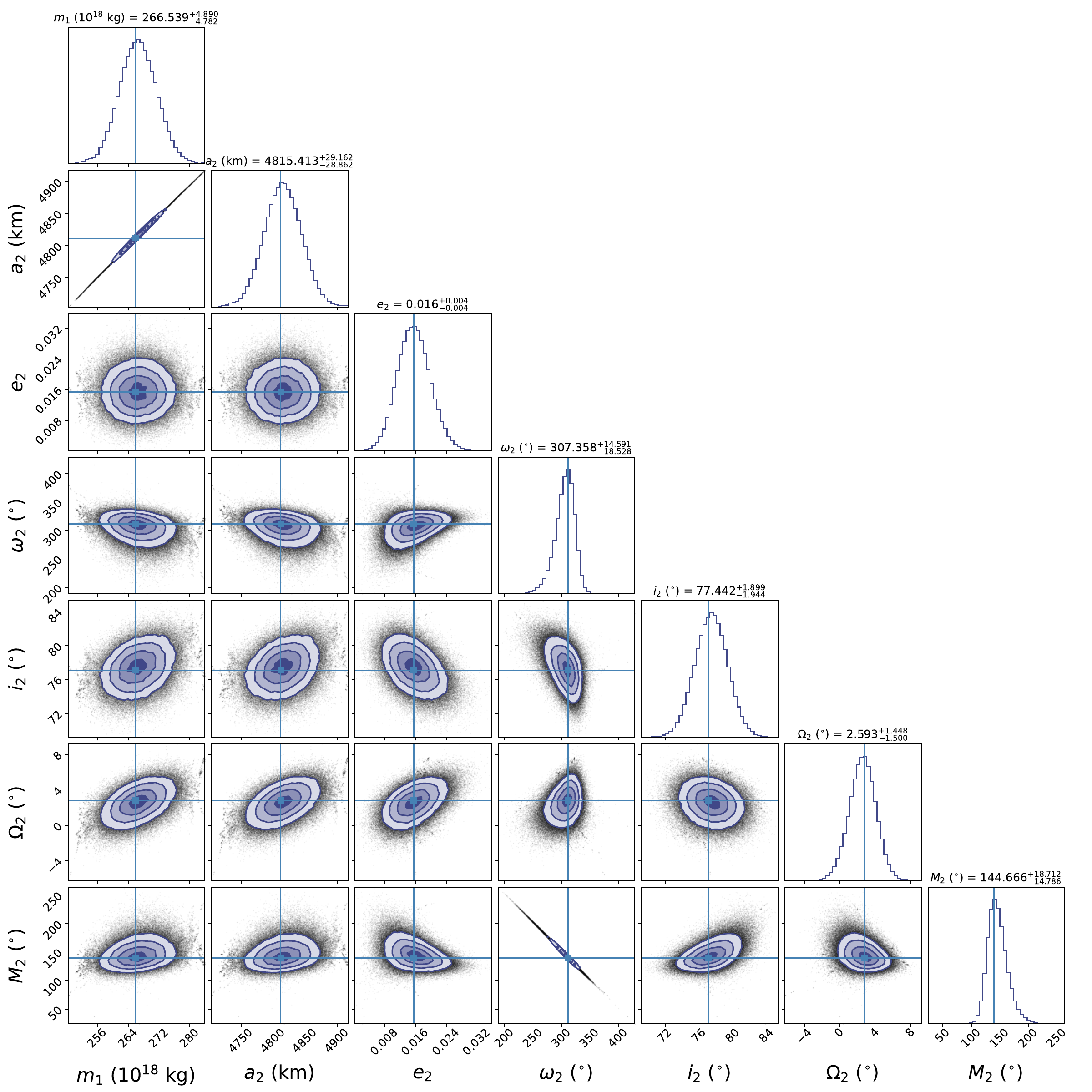}
    \caption{Corner plot for our Keplerian orbit fit of Varda-Ilmarë. Horizontal/vertical lines show the values of the best fit parameter set. All angles are referenced to the J2000 ecliptic frame.}
    \label{fig:corner_orbit}
\end{figure*}

In Figures \ref{fig:residual} and \ref{fig:corner_orbit}, we show the orbit fit residuals and the orbit fit posterior (as a corner plot).

\section{Occultation Chords}
Below, in Table \ref{tab:chords}, we list all occultation chords used in our shape fitting.

\FloatBarrier

\startlongtable
\begin{deluxetable*}{ccccc}
\tabletypesize{\footnotesize}
\tablecaption{Occultation Chords}
\tablehead{
Site Name       & Longitude    & Immersion time         & Emersion time          & Reference \\
                & Latitude     &                        &                        &           \\
                & Altitude (m) &                        &                        &           
                }
\startdata
\multicolumn{5}{l}{\textbf{\uk, 2014-11-15}} \\
\hline
                & $-$119 45 53.0 &                          &                          &                              \\
Reno            & +39 23 28.5   & 10:19:34.10 $\pm$ 0.70   & 10:19:46.64 $\pm$ 0.65   & \citet{benedetti2016results} \\
                & 1470         &                          &                          &                              \\
\hline
                & $-$119 47 46.8 &                          &                          &                              \\
Carson City (B) & +39 11 08.2   & 10:19:32.20 $\pm$ 0.50   & 10:19:47.30 $\pm$ 0.41   & \citet{benedetti2016results} \\
                & 1548.1       &                          &                          &                              \\
\hline
                & $-$119 33 31.4 &                          &                          &                              \\
Carson City (S) & +39 16 26.5   & 10:19:30.60 $\pm$ 0.60   & 10:19:46.30 $\pm$ 0.42   & \citet{benedetti2016results} \\
                & 1332.6       &                          &                          &                              \\
\hline
                & $-$119 40 20.3 &                          &                          &                              \\
Garderville     & +38 53 23.5   & 10:19:29.71 $\pm$ 0.46   & 10:19:48.07 $\pm$ 0.30   & \citet{benedetti2016results} \\
                & 1534.9       &                          &                          &                              \\
\hline
                & $-$119 09 39.0 &                          &                          &                              \\
Yerington       & +38 59 28.3   & 10:19:29.25 $\pm$ 0.41   & 10:19:46.00 $\pm$ 0.35   & \citet{benedetti2016results} \\
                & 1342.7       &                          &                          &                              \\
\hline
                & $-$117 14 06.7 &                          &                          &                              \\
Tonopah         & +38 05 22.1   & 10:19:17.6 $\pm$ 0.55    & 10:19:42.45 $\pm$ 0.30   & \citet{benedetti2016results} \\
                & 1838.7       &                          &                          &                              \\
\hline
                & $-$88 11 46.4  &                          &                          &                              \\
Urbana          & +40 05 12.5   & 10:18:03.61 $\pm$ 0.90   & 10:18:27.60 $\pm$ 0.90   & \citet{benedetti2016results} \\
                & 227          &                          &                          &                              \\
\hline
                & $-$119 24 45.0 &                          &                          &                              \\
Adler Springs   & +37 04 13.5   & 10:19:24.356 $\pm$ 0.159 & 10:19:50.249 $\pm$ 0.159 & \citet{schindler2017results} \\
                & 1405         &                          &                          &                              \\
\hline
\multicolumn{5}{l}{\textbf{Varda, 2018-09-10}} \\
\hline
& $-$114 35 48.9 &                          &                          &                              \\
MHV             & +35 01 54.1   & 03:32:57.10 $\pm$ 0.60   & 03:33:58.83 $\pm$ 0.60   & \citet{souami2020multi}      \\
                & 184          &                          &                          &                              \\
\hline
                & $-$114 28 13.1 &                          &                          &                              \\
TWF             & +42 35 01.9   & 03:33:56.99 $\pm$ 0.76   & 03:35:05.27 $\pm$ 0.30   & \citet{souami2020multi}      \\
                & 1133         &                          &                          &                              \\
\hline
                & $-$114 26 10.5 &                          &                          &                              \\
YMA             & +32 39 34.0   & 03:34:17.10 $\pm$ 0.60   & 03:35:26.39 $\pm$ 0.59   & \citet{souami2020multi}      \\
                & 97           &                          &                          &                              \\
\hline
                & $-$111 57 07.9 &                          &                          &                              \\
CAR             & +33 48 42.9   & 03:34:12.39 $\pm$ 0.47   & 03:35:19.52 $\pm$ 0.47   & \citet{souami2020multi}      \\
                & 654          &                          &                          &                              \\
\hline
                & $-$111 21 00.6 &                          &                          &                              \\
FLO             & +33 00 54.3   & 03:34:23.30 $\pm$ 0.13   & 03:35:26.75 $\pm$ 0.15   & \citet{souami2020multi}      \\
                & 484          &                          &                          &                             \\
\hline
\multicolumn{5}{l}{\textbf{Haumea, 2017-01-21}} \\
\hline
                                   & +20 14 02.1   &                          &                          &                              \\
Skalnate Pleso Observatory         & +49 11 21.8   & 03:08:26.79 $\pm$ 0.96   & 03:10:24.56 $\pm$ 0.8    & \citet{ortiz2017size}        \\
                                   & 1826         &                          &                          &                              \\
\hline
                                   & +19 53 41.5   &                          &                          &                              \\
Konkoly Observatory (1.0 m)               & +47 55 01.6   & 03:08:20.3 $\pm$ 0.2     & 03:10:17.39 $\pm$ 0.07   & \citet{ortiz2017size}        \\
                                   & 935          &                          &                          &                              \\
\hline
                                   & +19 53 41.5   &                          &                          &                              \\
Konkoly Observatory (0.6 m)               & +47 55 01.6   & 03:08:19.5 $\pm$ 0.8     & 03:10:16.4 $\pm$ 1.3     & \citet{ortiz2017size}        \\
                                   & 935          &                          &                          &                              \\
\hline
                                   & +14 46 53.3   &                          &                          &                              \\
Ondrejov Observatory               & +49 54 32.6   & 03:08:29.2 $\pm$ 0.8     & 03:10:12.2 $\pm$ 0.8     & \citet{ortiz2017size}        \\
                                   & 526          &                          &                          &                              \\
\hline
                                   & +12 00 44.0   &                          &                          &                              \\
Wendelstein Observatory (2.0 m)           & +47 42 13.6   & 03:08:27.9 $\pm$ 2.8     & 03:09:34.1 $\pm$ 0.5     & \citet{ortiz2017size}        \\
                                   & 1838         &                          &                          &                              \\
\hline
                                   & +12 00 44.0   &                          &                          &                              \\
Wendelstein Observatory (0.4 m)           & +47 42 13.6   & 03:08:18.8 $\pm$ 6       & 03:09:38.6 $\pm$ 6       & \citet{ortiz2017size}        \\
                                   & 1838         &                          &                          &                              \\
\hline
                                   & +11 36 25.2   &                          &                          &                              \\
Bavarian Public Observatory        & +48 07 19.2   & 03:08:30.0 $\pm$ 3.3     & 03:09:30.0 $\pm$ 4.9     & \citet{ortiz2017size}        \\
                                   & 538          &                          &                          &                              \\
\hline
                                   & +11 34 08.4   &                          &                          &                              \\
Asiago Observatory                 & +45 50 54.9   & 03:08:20.17 $\pm$ 0.08   & 03:09:13.27 $\pm$ 1.5    & \citet{ortiz2017size}        \\
                                   & 1376         &                          &                          &                              \\
\hline
                                   & +10 48 14.0   &                          &                          &                              \\
San Marcello Pistoiese Observatory & +44 03 51.0   & 03:08:22.9 $\pm$ 0.9     & 03:08:42.8 $\pm$ 0.9     & \citet{ortiz2017size}        \\
                                   & 965          &                          &                          &                              \\
\hline
                                   & +10 43 01.2   &                          &                          &                              \\
Lajatico Astronomical Centre       & +43 25 44.7   & 03:08:19.9 $\pm$ 1.4     & 03:08:34.3 $\pm$ 1.4     & \citet{ortiz2017size}        \\
                                   & 433          &                          &                          &                              \\
\hline
                                   & +10 30 53.8   &                          &                          &                              \\
Mount Agliale Observatory          & +43 59 43.1   & \nodata                  & \nodata                  & \citet{ortiz2017size}        \\
                                   & 758          &                          &                          &                             \\
\enddata
\tablecomments{Longitude is positive to the east, latitude is positive north. Occultation chords with no immersion or emersion times are close negative chords which are used to filter out two-dimensional ellipses that intersect it.}
\label{tab:chords}
\end{deluxetable*}

\end{document}